\documentclass[twocolumn,showpacs,superscriptaddress]{revtex4-2}

\usepackage[utf8]{inputenc}
\usepackage{graphicx}
\usepackage{epstopdf,epsfig}
\usepackage{amsmath}
\usepackage{amssymb}
\usepackage{amsbsy,amsxtra}
\usepackage{amsfonts}
\usepackage{braket}
\usepackage{dcolumn}   					% Needed for some tables
\usepackage{bm}        						% For math
\usepackage{amssymb}   					% For math
\usepackage[usenames]{color}
\usepackage{natbib}
\usepackage{float}
\usepackage{gensymb}
\usepackage{hyperref}
\usepackage{subfigure}
\usepackage{lipsum}
\usepackage{xcolor}

\begin{document}

\title{Resistive Switching Acceleration induced by Thermal Confinement}

\author{Alexandros Sarantopoulos}
 \email{a.sarantopoulos@fz-juelich.de}
  \affiliation{Peter Gruenberg Institute (PGI-7), Forschungszentrum Juelich GmbH and JARA-FIT, 52425 Juelich, Germany}

\author{Kristof Lange}
   \affiliation{IWE2, RWTH Aachen University, Aachen 52074, Germany}

\author{Francisco Rivadulla}
\affiliation{Centro de Investigaci\'{o}n en Qu\'{i}mica Biol\'{o}gica y Materiales Moleculares (CIQUS), Universidad de Santiago de Compostela, 15782 Santiago de Compostela, Spain}

\author{Stephan Menzel}
   \affiliation{IWE2, RWTH Aachen University, Aachen 52074, Germany}

\author{Regina Dittmann}
 \email{a.sarantopoulos@fz-juelich.de}
  \affiliation{Peter Gruenberg Institute (PGI-7), Forschungszentrum Juelich GmbH and JARA-FIT, 52425 Juelich, Germany}

\begin{abstract}

Enhancing the switching speed of oxide-based memristive devices at a low voltage level is crucial for their use as non-volatile memory and their integration into emerging computing paradigms such as neuromorphic computing. Efforts to accelerate the switching speed often result in an energy trade-off, leading to an increase of the minimum working voltage. In our study, we present an innovative solution: the introduction of a low thermal conductivity layer placed within the active electrode, which impedes the dissipation of heat generated during the switching process. The result is a notable acceleration in the switching speed of the memristive model system SrTiO$_{3}$ by a remarkable factor of $10^{3}$, while preserving the integrity of the switching layer and the interfaces with the electrodes, rendering it adaptable to various filamentary memristive systems. The incorporation of HfO$_{2}$ or TaO$_{x}$ as heat-blocking layers not only streamlines the fabrication process, but also ensures compatibility with complementary metal-oxide-semiconductor technology.

\end{abstract}


\maketitle

%%%%%%%%%%%%%%
%text
%%%%%%%%%%%%%%

\noindent\textbf{Introduction}

Memristive devices based on the valence change mechanism (VCM) have exhibited substantial potential not only in the domain of digital emerging memories but also in analog neuromorphic and in-memory computing applications \cite{waser2007nanoionics,dittmann2021nanoionic,yang2013memristive,borghetti2010memristive,adam2018challenges,covi2016analog,mehonic2022brain,christensen20222022,bagdzevicius2022interface,wang2021memristive}. Considering these applications, many endeavors have been directed towards enhancing the performance of memristive devices, with a particular focus on increasing the switching speed and minimizing the switching voltages.

Numerous studies have delved into the physical mechanisms behind the switching speed of memristive devices. These investigations have successfully showcased impressive switching speeds, often reaching the nano- and picosecond range \cite{pickett2009switching,torrezan2011sub,nishi2013origin,bottger2020picosecond,csontos2022picosecond}. Specifically, the interplay of both field and temperature acceleration results in a markedly non-linear relationship between switching speed and switching voltage \cite{menzel2011origin}. The increase in switching speed is at the expense of the voltage; for one of the fastest filamentary resistive switching systems (Ta$_{2}$O$_{5}$-based memristive device), a switching voltage of $\geq$ 2 V is required to switch the device with a 10 ns pulse, while switching the device with picosecond pulses, the necessary voltage is above 9 V \cite{bottger2020picosecond}. Hence, developing strategies to boost switching speed at a designated low switching voltage is imperative. This not only aids in minimizing power consumption but also aligns with the voltage requirements of highly scaled complementary metal-oxide-semiconductor (CMOS) transistors, essential for the CMOS co-integration of memristors.

Commonly, the most relevant parameters affecting the switching speed are besides the energy barrier for oxygen motion within the switching layer, the filament size and the oxygen exchange at the interface to the oxidizable electrode \cite{schroeder2010voltage,ielmini2012evidence,fleck2014interrelation,marchewka2016nanoionic}. In addition, for eightwise resistive switching devices, the oxygen exchange reaction with the Pt top electrode is often the rate determining step of the switching process, dictating the switching speed of the device \cite{cooper2017anomalous,siegel2021trade,dittmann2012scaling}. The filament size is to some extent an inherent property of the switching material, although external factors (e.g. the current compliance) significantly influence the filament dimensions \cite{ninomiya2013conductive}. Smaller filaments cause a stronger confinement of the current, amplifying Joule heating and consequently contributing to an increased switching speed \cite{menzel2011origin,fleck2014interrelation,west2023thermal,fleck2015set}.

Joule heating generated in the course of the switching process, increases the diffusion of oxygen vacancies in filamentary ReRAM devices, which leads to a substantial acceleration of the switching speed \cite{menzel2011origin,fleck2015set,menzel2015physics,huang2014analysis,russo2007conductive}. In order to minimize the heat losses through the thermally conductive metal electrode, the use of an electrode material with low thermal conductivity  might enable further switching speed acceleration. Although it is common and well accepted that thermal management is crucial for phase change materials \cite{khan2021ultralow,aryana2020thermal,bozorg2011temperature,harnsoongnoen2009confined,reifenberg2009thermal,xiong2011low,ahn2015energy}, for valence change memory devices this strategy has not been exploited so far, in terms of switching speed and operation voltages. 

In attempts to take advantage of a thermal enhancement layer, the introduction of low thermal conductivity layers at the oxide/metal interface resulted in a larger window between the high resistive state and the low resistive state (ON/OFF ratio) and in improved multilevel switching \cite{west2023thermal,lv2010improvement,liu2012improvement,wu2017improving,islam2022improved}. Nonetheless, a comprehensive physical explanation elucidating the influence of the thermal enhancement layer on device performance was absent. Moreover, directly interfacing the thermal enhancement layer with the memristive material, will significantly influence interface reactions, resulting in changes of the oxygen vacancy concentration and the oxygen exchange reaction with the electrodes. Therefore, modifying the switching layer and interfaces to improve switching speed may have a negative impact on other relevant properties, such as resistance range and programmed state retention \cite{siegel2021trade}. In this work, the devices are based on SrTiO$_{3}$ (STO), due to its very well known defect chemistry, ion diffusion mechanisms, and its wide use as a model system in VCM resistive switching, both in experiments and modeling \cite{,menzel2011origin,marchewka2016nanoionic,dittmann2012scaling,waser1991bulk,de2012behavior,schie2014simulation}.

It is important to note that modifying the materials and geometry of the device can have a detrimental effect on the overall performance. In particular, the interplay between thermal and electrical conductivity is crucial to our approach, as the two properties are closely intertwined. Ideally, a material with high electrical conductivity and very low thermal conductivity would be able to act as a thermal barrier layer without sacrificing electrical conductivity, thus providing resistive switching acceleration and a reduction in operating voltages. Since it is difficult to have both properties in CMOS-compatible materials as an electrode replacement, the combination of materials, dimensions and overall implementation must be such that the trade-off between thermal acceleration and increased resistance has a positive sign and the overall result is beneficial to device performance.

In this context, we present a novel approach for thermal management of VCM devices, which provides a solution for this trade-off. By incorporating a heat blocking layer (HfO$_{2}$ or TaO$_{x}$) positioned within the active electrode of our STO model system VCM devices, we have been able to achieve a remarkable acceleration in switching speed, reaching up to $10^{3}$ times faster in terms of time. In addition, this approach enables an approximate 30\% reduction in the switching voltage required to sustain the switching speed. This not only improves the energy efficiency of ReRAM cells, but also positions them well for integration with low-voltage transistor technology. The choice of HfO$_{2}$ or TaO$_{x}$ as heat barrier materials facilitates the implementation of our approach due to the wide availability of the materials and the simplicity of their fabrication process.

\vspace{0.1cm}\noindent\textbf{Results}

\begin{figure*}[t!]
		\centering
     \includegraphics[width=0.90\textwidth]{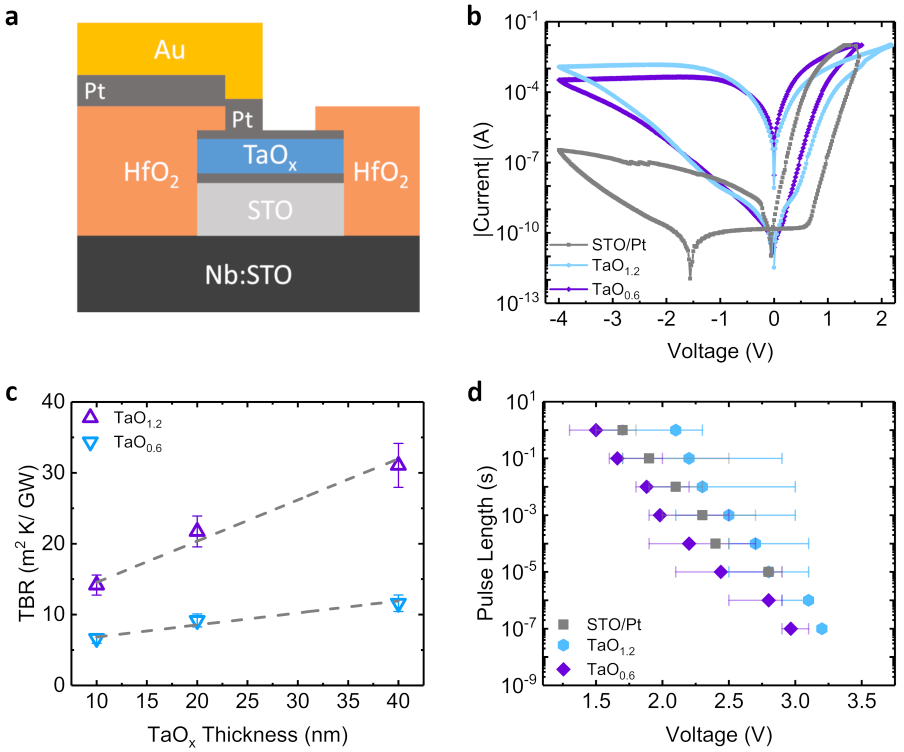}
	\vspace{-0px}\caption{\footnotesize{\textbf{TaO$_{x}$ heat blocking interlayer.} (a) Device layout sketch showing the TaO$_{x}$ layer placed inside the Pt active electrode. (b) I-V sweeps for the devices of the different samples, the reference STO/Pt stack (gray) and the devices with the TaO$_{x}$ interlayers of different oxygen content. (c) Thermal boundary resistance values for different thicknesses of the TaO$_{x}$ interlayers, in the two different oxygen concentrations. The dashed gray lines are the linear fit used to extract the thermal conductivity (slope) and the thermal resistance (intercept) of the thin films. (d) SET speed measurements for the reference sample STO/Pt and the devices with TaO$_{x}$ interlayers.}}
	\label{fig:TaOx}
 \end{figure*}

Our first approach to the thermal engineering of our devices is by introducing a heat-blocking layer of TaO$_{x}$ inside the Pt active electrode in a stack sequence of Nb:STO/STO(15 nm)/Pt(10nm)/TaO$_{x}$(30nm)/Pt(20nm) (Figure \ref{fig:TaOx}a). TaO$_{x}$ is a widely used material for resistive switching, whose electrical and thermal conductivity can be tuned by its oxygen stoichiometry \cite{landon2015thermal}, which allows us to explore the trade-off between low thermal conductivity (efficient heat-blocking) and low electrical conductivity (increased series resistance in the active electrode). We employed two different oxygen stoichiometries: TaO$_{1.2}$ and TaO$_{0.6}$, with electrical conductivities of $1.4\times10^{4}$ S/m and $3.75\times10^{5}$ S/m, respectively (see Table \ref{table:properties} and Supporting Figure 1). 

Due to the gradual switching nature of STO-based memristive devices, there is no clear forming and switching step during the I-V sweeps. For that reason, we define as forming and DC-switching voltage the necessary voltage to reach the selected current compliance (CC), while SET voltage will refer to the voltage pulse that invokes a change of the resistance of the devices $R_{HRS}/R_{LRS}\geq$10. The reference STO/Pt and STO/Pt/TaO$_{x}$/Pt devices exhibit eightwise switching typical for STO-based ReRAM devices (Figure \ref{fig:TaOx}b) \cite{cooper2017anomalous,heisig2022chemical,muenstermann2010coexistence,baeumer2016quantifying}. During forming and DC-switching the CC was kept at 10 mA and the devices present very stable behavior after multiple cycles. The forming and DC-switching voltage is similar between the reference STO/Pt and the TaO$_{0.6}$ devices, however, the devices with the TaO$_{1.2}$ interlayer exhibit increased values (see Supporting Figure 3) due to the voltage drop at the STO, caused by the reduced electrical conductivity of the TaO$_{1.2}$ layer.

To probe the effect of TaO$_{x}$ on the heat dissipation across the metal electrode, we have measured the thermal conductivity using frequency domain thermoreflectance (FDTR) \cite{schmidt2009frequency,malen2011fdtr}. The results of the thermal boundary resistance (TBR) across the Pt interface with the STO are presented in Figure \ref{fig:TaOx}c. For the fittings, the TaO$_{x}$ interlayer was treated as thermal interface between Pt and STO. We have investigated TaO$_{x}$ films with different thickness, to improve the accuracy of the thermal analysis (for more details see Table \ref{table:properties} and Supporting Figures 4 - 7). The reference sample of STO/Pt presents the lowest TBR $= 4.83 \pm 11\%$ m$^{2}$K/GW, while the TBR increases linearly with the thickness of the TaO$_{x}$ barrier, and it is higher for TaO$_{1.2}$ compared to the more electrically conductive TaO$_{0.6}$. The thermal conductivity of TaO$_{x}$ can be estimated from the linear fit of the plot in Figure \ref{fig:TaOx}c and the thickness of the oxide layer, resulting in $\kappa_{TaO_{1.2}} = 1.2$ W/m K and $\kappa_{TaO_{0.6}} = 5.9$ W/m K. The obtained values are very similar to previous reports of TaO$_{x}$ thin films, supporting our analysis \cite{landon2015thermal}.

The effect of the thermal barrier in the SET speed of the samples is summarized in Figure \ref{fig:TaOx}d. The SET speed for the reference sample of STO/Pt is in good agreement with previous studies, namely, we observe a strongly non-linear decrease of the required SET pulse length with the SET voltage height \cite{menzel2011origin,siegel2021trade}. Regarding the STO/Pt/TaO$_{x}$/Pt devices, the two different oxygen concentrations of the oxygen deficient tantalum oxide exhibit very different behavior: the TaO$_{1.2}$ interlayer has a slower SET speed than the reference STO/Pt sample, for the whole range of voltages used in the measurements. On the other hand, TaO$_{0.6}$ devices exhibit accelerated SET speed for the whole range of voltages compared to the reference sample STO/Pt. The difference is at least one order of magnitude in terms of pulse length, or a voltage reduction of $\approx 15\%$ at given pulse lengths.

\begin{table}[ht]
\caption{Electrical and thermal properties} % title of Table
\centering % used for centering table
\begin{tabular}{c c c c} % centered columns (4 columns)
\hline\hline %inserts double horizontal lines
Sample & $\sigma$ (S/m) & $\kappa$ (W/m K) & TBR (m$^{2}$ K/GW) \\ [0.5ex] % inserts table
%heading
\hline % inserts single horizontal line
STO/Pt & 6.2$\times 10^{6}$ & 44.5 & 4.8 \\ % inserting body of the table
%\hline
TaO$_{1.2}$  & 1.4$\times 10^{4}$ & 1.2 & 26.1 \\
%\hline
TaO$_{0.6}$  & 3.75$\times 10^{5}$ & 5.9 & 10.2 \\
%\hline
HfO$_{2}$ (1 nm) & - & 0.53 & 18.8 \\
%\hline
HfO$_{2}$ (2 nm) & - & 0.53 & 20.5 \\
%\hline
HfO$_{2}$ (3 nm) & - & 0.53 & 22.1 \\ [1ex] % [1ex] adds vertical space\hline %inserts single line
%\hline
\end{tabular}
\label{table:properties} % is used to refer this table in the text
\end{table}

\begin{figure*}[t!]
		\centering
     \includegraphics[width=0.90\textwidth]{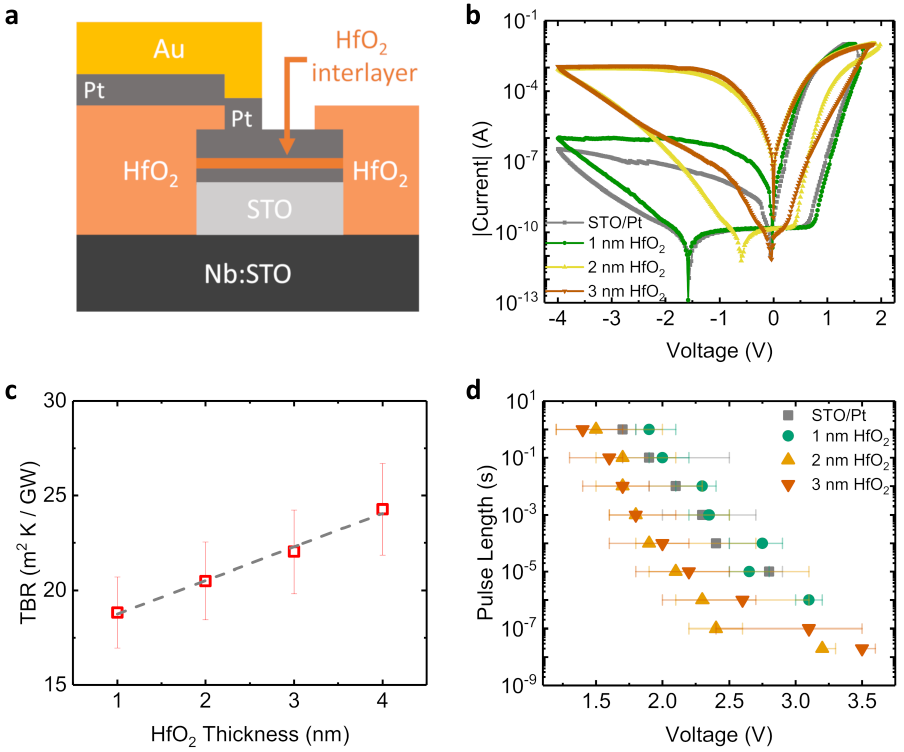}
	\vspace{-0px}\caption{\footnotesize{\textbf{HfO$_{2}$ heat blocking interlayer.} (a) Device layout sketch with the insertion of the HfO$_{2}$ interlayer inside the Pt active electrode. (b) I-V sweeps for the devices of the different thickness HfO$_{2}$ interlayers inside the Pt active electrode. (c) Thermal boundary resistance values for different thicknesses of the HfO$_{2}$ interlayers ranging from 1 to 4 nm thickness. The dashed gray line is the linear fit used to extract the thermal conductivity (slope) and the thermal resistance (intercept) of the thin films. (d) SET speed measurements for the reference sample STO/Pt and the three different thicknesses of the HfO$_{2}$ interlayers.}}
	\label{fig:HfO2}
 \end{figure*}

In our second approach to thermal engineering our devices, ultra-thin layers of electrically insulating HfO$_{2}$ were placed inside the Pt active electrode, as depicted in Figure \ref{fig:HfO2}a. The stack sequence is Nb:STO/STO(15 nm)/Pt(10nm)/HfO$_{2}$(1-3nm)/Pt(20nm). HfO$_{2}$ exhibits a very low thermal conductivity, making this a promising candidate as a heat blocking barrier. Since the HfO$_{2}$ interlayers are electrically insulating, their thickness has been kept sufficiently small to avoid an increased series resistance inside the Pt electrode, and make sure that the HfO$_{2}$ interlayer does not contribute to resistive switching. To confirm this point, additional control samples of Pt/HfO$_{2}$/Pt were prepared and measured in the same range of voltage and current as the original devices, showing no signs of distinguishable HRS and LRS (see Supporting Figure 2). The I-V sweeps in Figure \ref{fig:HfO2}b present eightwise resistive switching mode. The reference STO/Pt and 1 nm HfO$_{2}$ interlayer devices exhibit similar forming and DC-switching voltage, however the devices with 2 and 3 nm HfO$_{2}$ exhibit increased forming voltage and slightly higher DC-switching voltage (see Supporting Figure 3) due to the increased series resistance of the thicker HfO$_{2}$ interlayers.

For the analysis of the heat blocking efficiency of the HfO$_{2}$ interlayers, we measured the thermal resistances with FDTR. As discussed before, to improve the accuracy of the analysis, we treated the Pt/HfO$_{2}$/Pt as a single thermal resistance between the metal transducer and the STO film, in our fittings (for more details see Table \ref{table:properties} and Supporting Figures 4 - 7). The results of the TBR of these interfaces are presented in Figure \ref{fig:HfO2}c. Introducing just 1 nm thick interlayer of HfO$_{2}$ increases the TBR substantially ($18.8\pm 10\%$ m$^{2}$K/GW) with respect to STO/Pt in the reference sample ($4.83\pm 11\%$ m$^{2}$K/GW). The TBR increases linearly with the thickness of HfO$_{2}$, reaching $24.2\pm 11\%$ m$^{2}$K/GW for the 4 nm interlayer. Using the values for the thermal boundary resistance and the thickness extrapolation approach \cite{sarantopoulos2020reduction,lee1997heat,scott2018thermal}, we estimated a thermal conductivity of the HfO$_{2}$ layer $\kappa = 0.54 \pm10\%$ W/m K, in very good agreement with the literature \cite{scott2018thermal,panzer2009thermal}.  

The results for the SET speed measurements are plotted in Figure \ref{fig:HfO2}d: the devices with 1 nm-thick HfO$_{2}$ barrier exhibit a similar speed as the control sample, slightly slower at lower voltages and slightly faster at higher voltages. Increasing the thickness of the HfO$_{2}$ interlayer produces a significant improvement, even at lower voltages. With increasing pulse height, the difference between reference and HfO$_{2}$ interlayer devices becomes more pronounced. The results show an acceleration of the SET process for a given pulse length, of x10$^{3}$, reducing the SET time from 100 $\mu$s to 100 ns in the range of 2.3 - 3.2 V. Also, the results may be described from an energy efficiency point of view, noting that the same pulse length reduces the required voltage pulse by $\approx 20 \%$ at the lower voltages, and up to $\approx 30\%$ at higher voltages.

\begin{figure*}[t!]
		\centering
     \includegraphics[width=0.97\textwidth]{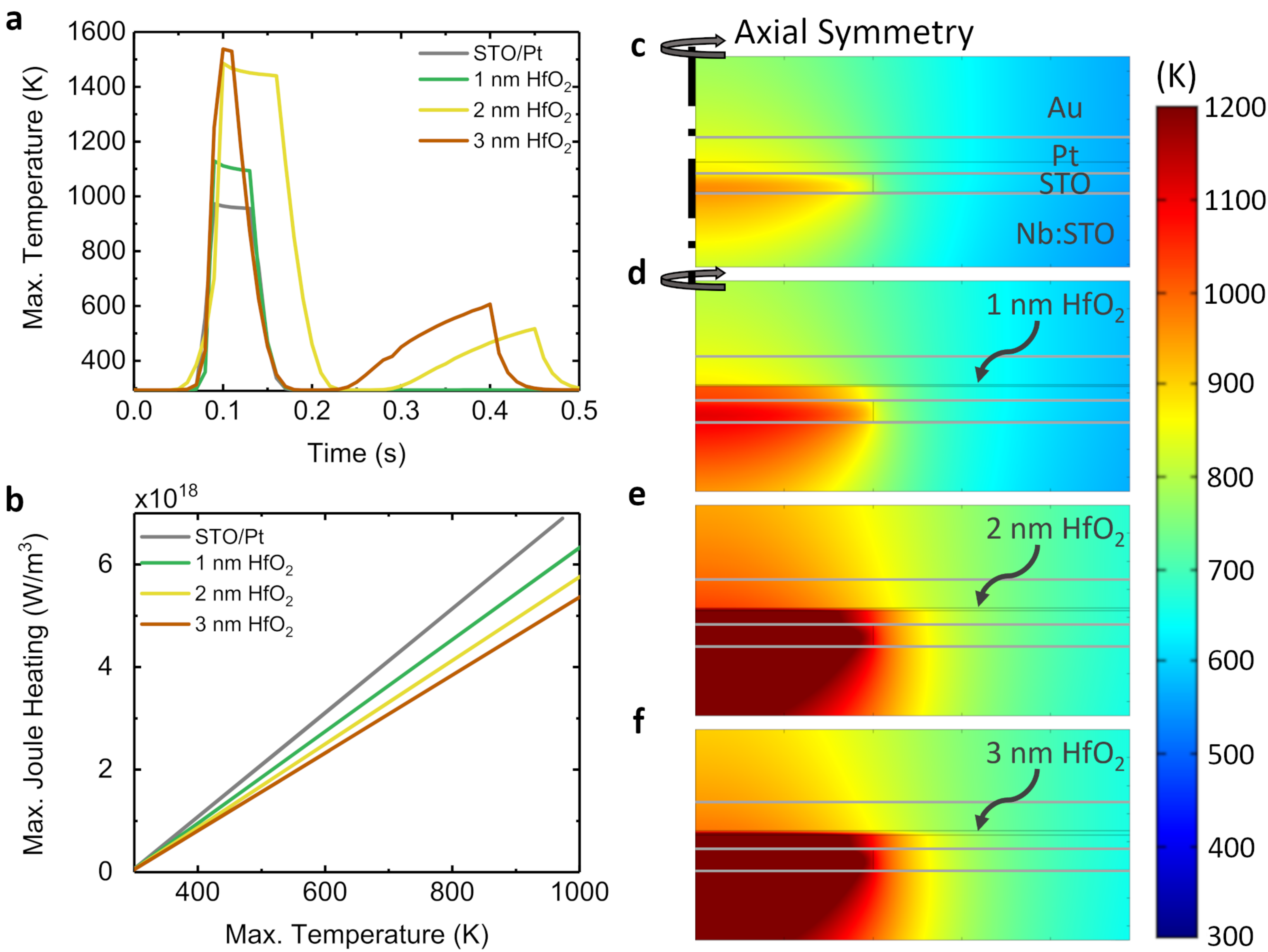}
	\vspace{-0px}\caption{\footnotesize{\textbf{Calculations of local temperature.} (a) Maximum temperature reached for each sample calculated by the currents of the I-V sweeps as input, as a function of time. (b) Heating efficiency of each sample based on the necessary Joule heating needed to reach the maximum temperature for each sample. 2D temperature profiles at $t = 0.1$ s for the reference STO/Pt configuration without interlayer (c) and the configurations with HfO$_{2}$ interlayer thicknesses of 1 nm (d), 2 nm (e), and 3 nm (f), respectively. A vertical boundary line in the STO layer marks the outer edge of the filament. The additional horizontal line in the Pt layer reflects the two-step platinum evaporation, in order to ensure that the observed behavior is due to the insertion of the interlayers. The arrows indicate the position of the interlayers inside the Pt electrode.}}
	\label{fig:simulations}
 \end{figure*}

In order to quantify the effect of the HfO$_{2}$ barrier on the dissipation of heat across the Pt layer we simulated by finite element method (FEM) simulations, using the device structure shown in Supporting Figure 8. This structure is based on the device layout sketched in Figure \ref{fig:HfO2}a. The only exterior electrical as well as thermal contacts are located at the outer edge of the Au layer and at the bottom of the Nb:STO layer. In Supporting Table 2, the material parameters used for simulation are listed. Taking as input the experimental data from the I-V sweeps, we were able to calculate the maximum temperature reached at the filament vicinity for all samples, depicted in Figure \ref{fig:simulations}a. The 2 nm HfO$_{2}$ interlayer sample exhibits a longer sustained temperature due to the slightly higher voltage that needed to be applied in order to reach the CC for the I-V sweep, compared to the other devices (see Figure \ref{fig:HfO2}b). Moreover, the heating efficiency of each sample has been calculated and plotted in Figure \ref{fig:simulations}b.  An overview of the local maxima for the temperature within the STO layer is presented in Fig. \ref{fig:simulations}c-f as 2D temperature profiles. The arrows indicate the position of the HfO$_{2}$ interlayers within the Pt electrode.

Figure \ref{fig:simulations}c-f shows 2D temperature profiles for the reference STO/Pt device and the three HfO$_{2}$ interlayer thicknesses, at $t = 0.1$ s when the overall highest temperatures are observed. The calculated temperature at the filament region within the STO layer increases with the HfO$_{2}$ interlayer insertion compared to the reference device, further increasing for the thicker interlayers. At the position of the interlayer (marked with arrows in Fig. \ref{fig:simulations}) the temperature exhibits an abrupt decrease at the upper part of the interlayer, highlighting the increased thermal resistance of the HfO$_{2}$ within the platinum electrode. For thicker interlayer, the temperature difference at the two sides of the interlayer becomes larger. As a consequence of the heat-blocking effect of the HfO$_{2}$ interlayer, the temperature at the filament region can be up to $\approx 500$ K higher for the thicker interlayers compared to the reference STO/Pt devices. 

Here we must note that the temperature calculations are very sensitive to the choice of the filament dimensions. We have chosen the filament radius as $r_{filament} = 200$ nm, based on previous reports on STO-based ReRAM devices \cite{baeumer2016quantifying,lenser2015formation}. The results on the local temperatures and the heating efficiency for other filament diameters are presented in Supporting Figure 9.

\vspace{0.1cm}\noindent\textbf{Discussion}

All the samples studied in this work exhibit similar I-V characteristics as have been reported in the literature for crystalline STO devices \cite{cooper2017anomalous,muenstermann2010coexistence,baeumer2016quantifying}. Here it is important to remind that the slightly increased DC-switching voltage for the devices with the heat-blocking layers refer to the voltages required to reach the 10 mA of CC during the I-V sweeps (Figures \ref{fig:TaOx}b and \ref{fig:HfO2}b), and not to the voltage pulses required to SET the devices with an ON/OFF $\geq$ 10 presented in Figures \ref{fig:TaOx}d and \ref{fig:HfO2}d. The forming voltage for almost all devices with embedded interlayers, except for the case of the high electrically conductive TaO$_{0.6}$ (see Supporting Figure 3) are slightly increased compared to the reference STO/Pt devices. This is due to the increased series resistance introduced by the interlayer, which leads to a voltage drop across the device, resulting at a lower voltage that is actually applied at the STO/Pt interface. Nevertheless, the DC-switching voltage is in the similar range for all the devices (see Supporting Figure 3). 

Although the dissipation of heat through the Nb:STO substrate is also important, the active role of the STO/Pt interface, with oxygen being stored/released in Pt during the SET/RESET process, makes the local temperature at this interface particularly relevant for the speed of this process. Moreover, the lower thermal conductivity of the Nb:STO substrate ($\kappa = 8.7 \pm 10\%$ W/m K) compared to the Pt electrode ($\kappa = 44.5 \pm 10\%$ W/m K) ensures that the heat dissipation through the Pt is much larger, making it our point of focus for the thermal confinement approach. The FDTR measurements have confirmed the heat blocking efficiency of the TaO$_{x}$ and HfO$_{2}$ interlayers, compared to the Pt electrode, reducing the heat dissipation along this channel, which in turn increases the local temperature and increases the mobility of oxygen vacancies during the SET process \cite{west2023thermal,wu2017improving,islam2022improved,rieck2021trade,hensling2018tailoring}. The increased local temperature is also confirmed by the calculated temperatures in Figure \ref{fig:simulations}a and c-f, where the temperature increase is $\approx 150$ K for the 1 nm-thick HfO$_{2}$ interlayer and about $\approx 500$ K for the 2 and 3 nm-thick interlayers. The accelerated ionic movement is confirmed by the results of the SET speed presented in Fig. \ref{fig:TaOx}d and \ref{fig:HfO2}d, where the SET speed is increased when the heat confinement effect is larger. Moreover, the samples with thicker HfO$_{2}$ interlayers exhibit a higher heating efficiency (see Figure \ref{fig:simulations}b), meaning that in order to reach the same temperature, less Joule heating is needed for the samples with thicker HfO$_{2}$ interlayers. This directly translates to reduced current levels in order to achieve the same temperatures, thus improving the energy efficiency of the devices.

On the other hand, the electrical resistance between the two Pt electrodes in the Pt/HfO$_{2}$/Pt samples is $\approx$ 50, 60, and 75 $\Omega$, for 1, 2, and 3 nm of HfO$_{2}$, respectively (see Supporting Figure 2). Considering the CC of 10 mA at higher voltages, the estimated voltage drop would correspond to 0.5, 0.6, and 0.75 V, respectively. This small increment in electrical resistance is overcompensated by the much larger increase in TBR, which goes from 18.8 m$^{2}$K/GW to 20.5 m$^{2}$K/GW, along this series. Thus, this analysis suggests that the positive effect of ionic acceleration due to a higher local temperature, compensates the negative impact on the electrical conductivity of the Pt active electrode after inserting the heat-blocking layer. This trade-off between ionic acceleration and increased series resistance is clearly visible in the results of the SET speed of the HfO$_{2}$ interlayer samples, compared to the reference STO/Pt devices (Figure \ref{fig:HfO2}d). The ionic acceleration for the 1 nm-thick HfO$_{2}$ interlayer is compensated by the increase in series resistance inside the Pt electrode. However, the positive effect of the thermally accelerated ionic motion is highlighted for the devices of the 2 and 3 nm-thick HfO$_{2}$ interlayer, achieving faster SET times up to 3 orders of magnitude compered to the devices of the STO/Pt reference sample.

In the case of the TaO$_{x}$ samples, we observe a similar behavior, with the TBR increasing linearly with increasing thickness. The different slopes in the linear increase of the TBR for the two different oxygen concentrations of the TaO$_{x}$ samples reflect the different thermal conductivity values obtained earlier. The TBR value for the more oxygen deficient TaO$_{0.6}$ is closer to the TBR value for the reference STO/Pt interface, due to its high electrical conductivity and different stoichiometry compared to its less oxygen deficient TaO$_{1.2}$. In the latter case, the TBR values are much higher, closer to the values obtained for the HfO$_{2}$ interlayers. This effect is also reflected in the results of the SET speed depicted in Figure \ref{fig:TaOx}d, where the more electrically and thermally resistive TaO$_{1.2}$ interlayer devices exhibit slower SET speeds for the whole range of voltages and times. On the other hand, the more electrically (and thermally) conductive TaO$_{0.6}$ interlayer devices show a constant acceleration of the SET speed of up to one order of magnitude in terms of time, compared to the STO/Pt devices. Here, it is important to point out that the TBR values of the TaO$_{x}$ interlayers are becoming comparable to the HfO$_{2}$ interlayer TBR values, when the thickness of TaO$_{x}$ is above 20 nm for the case of the less oxygen deficient TaO$_{1.2}$ and always lower for the case of TaO$_{0.6}$ even at 40 nm. The trade-off between thermal and electrical conductivity should be carefully taken into consideration and the thickness and stoichiometry of the heat blocking interlayers should be optimized accordingly.

In conclusion, introducing the heat-blocking interlayers has been successful with respect to the acceleration of the switching speed of the STO-based ReRAM devices, achieving up to 10$^{3}$ faster times. However, the switching speed is not the only crucial parameter in the device operation, where the operating voltage for the devices dictate the energy consumption of the device array in potential computing architectures. We have shown that by implementing the thermal confinement approach, the devices can be operated at the same switching speeds, but at $\approx$ 30$\%$ lower voltages. This reduction in energy cost can be even more appreciated when the energy consumption of the overall number of devices and operations are taken into account. This opens a window towards more efficient device operation with a strategy that is CMOS compatible, making our approach easily implemented in established processes of ReRAM device fabrication.

\vspace{0.2cm}\noindent\textbf{Methods}

\noindent\textbf{Sample \& Device Fabrication} The STO thin films were fabricated by pulsed laser deposition (PLD), grow epitaxially on Nb-doped ($0.5\%$wt) STO substrates, at $800\celsius$ and 0.1 mbar O$_{2}$ pressure, at 5 Hz. The growth process was monitored by reflection high-energy electron diffraction (RHEED) and the observation of oscillations confirmed the layer-by-layer growth. The STO films were intentionally grown as Sr-rich, which according to previous works, enhances the retention performance of the devices \cite{rieck2021trade}, but can also lead to slower SET speed \cite{siegel2021trade}. The surface morphology was examined by atomic force microscopy (AFM) images and the crystalline quality by x-ray diffraction (XRD) measurements.

Once the STO thin films were structurally characterized, 10 nm of Pt were thermally evaporated to form the Schottky barrier with the STO thin film. To minimize the sample-to-sample variability on the performance of the devices, we have used the same 10x10 mm STO thin film to prepare the different samples, by cutting it into four different pieces. The pieces were separated after the evaporation of the Pt top electrode, to ensure that the STO/Pt interface meets the quality standards and is undisturbed by possible side-effects during the cutting of the sample. In continuation, three samples underwent the HfO$_{2}$ interlayer sputtering deposition of 1, 2 and 3 nm thickness while one was kept as a reference sample. The TaO$_{x}$ interlayers were sputtered using a Ta target with background atmosphere of Ar mixed with $1\%$ O$_{2}$ gas, at different pressures. In the next step, an additional 20 nm of Pt was evaporated on all samples, so any different performance of the devices are attributed to the presence and thickness of the interlayers, when compared to the reference sample. The stoichiometry of the HfO$_{2}$ was confirmed by X-ray photoemission spectroscopy (XPS) in-situ after sputtering, while the stoichiometry for the TaO$_{x}$ interlayers was determined based on previous work by T. Heisig \emph{et al.}, from which the sputtering parameters where adapted \cite{heisig2022chemical}.

The patterning of the devices was performed by optical lithography in 3 different steps. First, the device area was patterned and then the rest of the Pt/Interlayer/Pt was etched away, together with the 15 nm of STO underneath. In the next step, the contact area of the subsequent top leads was defined and the rest of uncovered area was sputtered with $\approx$ 45 nm of HfO$_{2}$ to electrically insulate the bottom electrode (Nb:STO substrate) from the subsequently thermally evaporated top leads. As a last step, the top leads were patterned and thermally evaporated, 10 nm of Pt and 60 nm of Au. For the samples with the reduced tantalum oxide as a thermal barrier, the same procedure was followed. In order to avoid any further oxidation of the TaO$_{x}$ thermal barriers, an additional 10 nm of Pt were sputtered \emph{in-situ}. The rest of the lithography process was similar, as described before. 

The devices were formed with a positive voltage applied to the top electrode, while the bottom electrode (Nb:STO substrate) was grounded. After the forming step, the devices were switched multiple times, using voltages of $+2.5$ and $-4.0$ V for the SET and RESET process respectively. The forming and SET process were limited by a 10 mA current compliance. 

\noindent\textbf{Frequency Domain Thermoreflectance (FDTR)} The thermal conductivity of the different layers and their thermal boundary conduction were measured with the frequency domain thermoreflectance (FDTR) technique \cite{schmidt2009frequency,malen2011fdtr}. This is a non-contact \emph{pump-probe} optical technique based on the dependence of the reflectance of a metal transducer with its local temperature. In this technique, a sinusoidally modulated continuous wave (cw) pump laser ($\lambda = 488$ nm, $f = 0.5 - 5$ MHz) is directed on the sample surface, coated with a metallic transducer layer (Au: 40-60 nm), and produces a modulation of the surface temperature. The change of the thermoreflectance of the Au is monitored by the \textit{probe cw} laser. The effective thermal conductivity $\kappa_{eff}$ and the thermal boundary resistance (\textit{TBR}) between the different layers is calculated by the phase lag between the \textit{pump} and the \textit{probe} signals. 

To extract the thermal conductivity $\kappa$ and the thermal boundary resistance (TBR) from the FDTR phase-shit curves, we fitted them with a model wherein total energy conservation and energy transfer between layers are imposed by a transfer matrix, as explained elsewhere \cite{schmidt2009frequency}. More details of the technique and results on the thermal conductivity of thin films can be found in \cite{sarantopoulos2020reduction,sarantopoulos2018effect}. To reduce the number of fitting parameters, we first measured the thermal conductivity of the Nb:STO substrate ($\kappa = 8.7 \pm 10\%$ W/m K) and subsequently of the STO thin film ($\kappa = 5.4 \pm 10\%$ W/m K). The thermal conductivity of the Au and Pt layers were determined by measuring the electrical conductivity and using Wiedemann-Franz's law ($\kappa/\sigma = L T$), and then were further adjusted for optimal fitting of the data. 

\noindent\textbf{Electrical characterization \& Kinetics measurements}

The electrical characterization of the devices for the I-V sweeps were performed on a probe station where the top electrodes were contacted via tungsten whisker needle and the Nb:STO substrate (bottom electrode) was contacted through aluminum wire-bonding. As current source, a Keithley 2611A Source Meter was used. The typical voltage step was 30 mV/s with a waiting time of 5 ms between each step. The voltage was increased from 0 V to +3 V for the electroforming of the devices, followed by a sweep from 0 V to -4 V, to reset the devices to their high resistive state. Subsequent I-V sweeps were repeated from 0 V to +2.5 V to -4 V and back to 0 V in order to cycle the devices and ensure their stable behavior. The resistive state of the devices were determined by I-V sweeps from -0.3 V to + 0.3 V, and using a linear fit of the slope. The current compliance was set at 10 mA for all devices. The pulsed measurements were performed at a different probe station using tungsten whisker needles, as described before, and as a power source, a Keithley 4200, using a custom software written in a LUA environment.

The SET speed of the devices were measured according to the following protocol: the device is reset to a high resistive state (HRS) by applying pulses of defined length (time) and progressively increasing height (voltage), until a defined value is reached (600 M$\Omega$ $\pm 20\%$). The value of the device resistance is measured with a read-out (RO) pulse with height of $0.3$V and $60$ms length, before and after every reset pulse. Once the defined HRS is reached, the SET pulses begin, starting with the lowest height ($1.5$V) and applying pulses with increasing length of one order of magnitude (here we start with 20 ns instead of 10 ns due to the equipment limitation). As in the RESET process, before and after each SET pulse, the resistance is read-out at $0.3$V. If one of the pulses bring the device resistance to a value lower than the defined HRS, the RESET process starts again, preparing the device for the next SET pulse. When a SET pulse achieves a successful SET event, the rest of the pulse lengths for the same voltage are skipped and the process continues with for the next pulse height in the sequence. Here, we define as a successful SET event when $\frac{R_{HRS}}{R_{LRS}}>10$.

\begin{acknowledgments}

\noindent This work was supported by the DFG (German Science Foundation) within the collaborative research center SFB 917 ``Nanoswitches", by the Helmholtz Association Initiative and Networking Fund under Project No. SO-092 [Advanced Computing Architectures (ACA)] and the Federal Ministry of Education and Research (Project NEUROTEC, Grants No. 16ME0398K and No. 16ME0399).

\end{acknowledgments}

\noindent\textbf{Supplementary information}

Supplementary information can be found in a separate file.

\noindent\textbf{Author contributions}
\noindent A.S and R.D conceived and designed the experiments. K.L. and S.M. conceived and performed the simulations. A.S. prepared the samples and did all experiments and data analysis. F.R. contributed to the FDTR measurements and data analysis. A.S with R.D. wrote the manuscript with contributions from all authors.

\noindent\textbf{Competing interests}
\noindent The authors declare no competing interests.

\noindent\textbf{Data availability}
\noindent All data supporting the results of this study are available in the manuscript and the supplementary information. Additional data are available from the corresponding author upon request.

\bibliography{mybibSET}{}
\bibliographystyle{apsrev4-2}

\end{document}